# Extraordinary thermoelectric performance of ABaX compared to Bi$_2$Te$_3$

Enamul Haque

EH Solid State Physics Laboratory, Longaer, Gaffargaon-2233, Mymensingh, Bangladesh.

Email: enamul.phy15@yahoo.com, enamul@mailaps.org

**Abstract**

Thermoelectric materials can generate electricity directly utilizing heat and thus, they are considered to be eco-friendly energy resources. The thermoelectric efficiency at low temperatures is impractically small, except only a few bulk materials (Bi$_2$Te$_3$ and its alloys). Here, I predict two new thermoelectric materials, LiBaSb and NaBaBi, with excellent transport properties at low - medium temperature by using the first-principles method. The relatively low density of states near Fermi level, highly non-parabolic bands, and almost two times wider bandgap of NaBaBi lead to almost two times higher anisotropic power factor at 300K than that of Bi$_2$Te$_3$. On the other side, almost similar phonon density of states and anharmonicity of NaBaBi cause almost identical lattice thermal conductivity (but it is much higher in LiBaSb). These effects make it a superior thermoelectric material, with a predicted cross-plane (in-plane) ZT ~2 (~1) at 300 K for both n- and p-type carriers, even higher (~2.5 for p-type) at 350K. On the other hand, the isotropic maximum ZT of NaBaBi is ~1.2 and 1.6 at 350K for n and p-type carriers, respectively. However, LiBaSb is less suitable for low-temperature TE applications, because of its relatively wider bandgap and high lattice thermal conductivity.

In recent decades, researchers have been searching for new thermoelectric materials, because these materials are the clean energy resources and can converts waste heat into electricity. Some TE materials show the best efficiency at high temperatures while only a few materials exceptionally exhibit a large thermoelectric figure of merit at low temperatures. Generally, semiconducting materials have the best TE performance. At the room temperature region, the electrical conductivity of these materials is usually low, at the same time the lattice thermal conductivity shows opposite trends. Thus, there are only a few high-performance TE materials reported for low-temperature applications.

Bismuth telluride (Bi$_2$Te$_3$) and its alloys are the best known low-temperature TE materials and used commercially in cooling devices and thermoelectric generators. Especially, Bi$_2$Te$_3$, alloying with Sb$_2$Te$_3$ for p-type carrier and Bi$_2$Se$_3$ for the n-type carrier, exhibits the highest thermoelectric figure of merit (ZT) near the room temperature region [1]. The ZT is defined by the expression [2]

$$\text{ZT} = \frac{S^2 \sigma}{\kappa_{tot}} T$$

where S, σ, T, and $\kappa_{tot}$ stand for the Seebeck coefficient, electrical, absolute temperature, and total thermal conductivity (electronic plus phononic contribution to the thermal conductivity), respectively. The narrow bandgap, band degeneracy, and light effective mass induce high electrical conductivity in $Bi_2Te_3$ and its alloys [3]. The intense phonon scattering leads to high anharmonicity and hence low lattice thermal conductivity in these materials. To date, the reported maximum ZT of bulk $Bi_2Te_3$ and its alloys ($Bi_{0.5}Sb_{1.5}Te_3$ and $Bi_2Te_{2.7}Se_{0.3}$) is ~0.4 [3] and ~0.9 [4] at room temperature. Therefore, thermoelectric device performance will be dramatically increased by optimizing these parameters further or discovering a novel material with superior TE performance.

NaBaBi and LiBaSb are polar intermetallic compounds and were synthesized in 2004 [5] and 2001 [6], but, to date, crystal and electronic structure have been reported only. Sun et al. reported from the first-principle study that NaBaBi has a left-handed spin texture in the upper cone (for the band inversion between Bi-6p and Na-2s) like in a common trivial insulator $Bi_2Se_3$ [7]. Here I report the details of carrier and phonon transport properties of $Bi_2Te_3$, NaBaBi, and LiBaSb. Although both experimental and theoretical studies have been reported on $Bi_2Te_3$, I hereby repeated it to check my computational accuracy and compare the calculated parameters of these compounds.

Table I. Computed lattice parameters, elastic moduli, sound velocity, and Debye temperature along with available experimental data.

| Parameters | $Bi_2Te_3$ | | NaBaBi | | LiBaSb | |
|---|---|---|---|---|---|---|
| | Calc. | Exp. [8] | Calc. | Exp. [5] | Calc. | Exp. [6] |
| a (Å) | $a_R$=10.397 | 10.473 | 8.575 | 8.591 | 4.891 | 4.898 |
| c (Å) | $α_R$=24.342° | 24.159° | 5.107 | 5.091 | 8.990 | 9.014 |
| $c_{11}$ (GPa) | 73.7 | 74.4 [9] | 43.3 | - | 63.9 | - |
| $c_{12}$ (GPa) | - | - | 13.1 | | 11.6 | |
| $c_{13}$ (GPa) | 23.7 | 29.2 [9] | 17.2 | - | 16.4 | - |
| $c_{14}$ (GPa) | 13.1 | 15.4 [9] | - | - | - | - |
| $c_{33}$ (GPa) | 47.8 | 51.6 [9] | 53.2 | - | 42.3 | - |
| $c_{44}$ (GPa) | 28.2 | 29.2 [9] | 22.3 | - | 23.7 | - |
| $c_{66}$ (GPa) | 25.1 | 26.2 [9] | - | - | - | - |
| B (GPa) | 36.2 | 39.5 [9] | 25.8 | - | 28.5 | - |
| G (GPa) | 21.8 | - | 17.7 | - | 22.6 | - |
| $v_t$ (km s$^{-1}$) | 1.66 | - | 1.77 | - | 2.23 | - |
| $v_l$ (km s$^{-1}$) | 2.87 | - | 2.96 | - | 3.59 | - |
| $θ_D$ (km s$^{-1}$) | 169.9 | 164.9 [9] | 176.6 | - | 229.7 | - |

I performed full structural relaxation and electron-phonon calculations in Quantum Espresso (QE) code [10] by using generalized gradient approximation (GGA) [11] with the PBEsol [12] setting. For average electron-phonon dynamical matrix calculations, I used EPA code [13] based on energy bins and checked its sensitivity to the bins by EPA-MLS code [14]. As the Boltzmann transport equation (BTE) implemented in BoltzTraP [15] uses constant relaxation time approximation, the code was slightly modified to take into account the average e-ph matrix and hence calculate the carrier lifetime. The electronic parameters, such as eigenvalues, required for BoltzTraP, were obtained by using the FP-LAPW method implemented in Wien2K [16]. In the electronic structure calculations, I used an accurate (as GGA-PBEsol severely underestimates bandgap) mBJ potential [17] including the spin-orbit coupling (SOC) effect. The lattice thermal conductivities of these compounds were obtained by using the finite displacement method in Phono3py [18]. In this case, I created supercells and calculated the second and third-order interatomic force constants (IFCs) by QE. Please see the supporting information (SI) for the detailed explanations of these calculations.

Fig. S1 (top panel) shows the fully relaxed conventional and primitive unit cell of $Bi_2Te_3$. It forms a hexagonal crystal (space group $R\bar{3}m$, #166) and primitive rhombohedral unit cell [8]. In this structure, Bi-Te layers are bonded through strong covalent-ionic bonds, while tellurium layers are kept together by weak Vander-Walls forces. Unlike $Bi_2Te_3$, ABaX crystallize in a non-layered hexagonal structure (space group $P\bar{6}2m$ and $P6_3/mmc$, #189 [5], and #194 [6], respectively) as shown in the middle and bottom panel of the figure.

The computed lattice parameters, elastic constants, and related parameters are listed in Table I. The obtained lattice parameters show overall good agreement (deviations from experimental values are less than 1%). Moreover, the calculated elastic parameters and Debye temperature ($\theta_D$) of $Bi_2Te_3$ fairly agree with the reported experimental value. The slightly higher Debye temperature of NaBaBi compared to $Bi_2Te_3$ suggests its almost similar lattice thermal conductivity. In contrast, the $\theta_D$ of LiBaSb is much higher, indicating that the lattice thermal conductivity of LiBaSb would be much higher than that of $Bi_2Te_3$.

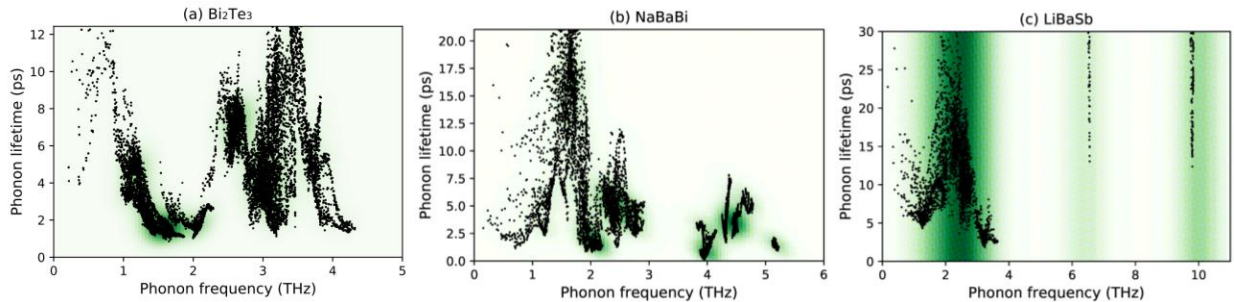

Fig. 1. Phonon lifetime of (a) $Bi_2Te_3$, (b) NaBaBi, and (c) LiBaSb calculated from anharmonic IFCs.

Phonon dispersion relations and phonon density of states of these compounds are shown in Fig. S3. Positive phonon energy ensures the dynamical stability and again, the computed phonon energy of $Bi_2Te_3$ shows overall good agreement with experimental values. In the former two compounds, Bi has the dominant contribution to the lower energy acoustic phonons, while in LiBaSb, Ba dominates in this region. The phonon lifetime and group velocity, as shown in Fig. 2 and Fig. S4 respectively, of NaBaBi, suggests the slight weak phonon scattering compared to $Bi_2Te_3$, but the scattering is much weaker in LiBaSb. The computed model Gruneisen parameter also shows the same trend, i.e., the modest phonon anharmonicity in NaBaBi and weak anharmonicity in LiBaSb. The weak anharmonicity in LiBaSb might be caused by Ba-induced acoustic phonons, because, unlike the former two compounds, Ba has the dominant contributions to the lower energy acoustic phonons. These features suggest that the amount of heat conduction by phonons in NaBaBi would be almost the same as that of $Bi_2Te_3$, but too different in LiBaSb.

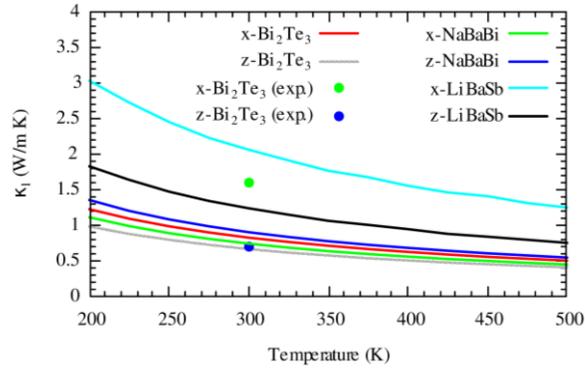

Fig. 2. Computed in-plane and cross-plane lattice thermal conductivity at different temperatures. The circles represent the experimental lattice thermal conductivity of $Bi_2Te_3$ taken from Ref. [3].

The computed lattice thermal conductivities ($\kappa_l$) provide the same trend, as shown in Fig. 3. Although the computed cross-plane lattice thermal conductivity ($\kappa_l$) of $Bi_2Te_3$ fairly agrees with the experimental value [3], it severely underestimates the in-plane $\kappa_l$ (about 48%). As the lattice parameters of $Bi_2Te_3$ are highly anisotropic, the DFT may not account for the rigid bonds correctly. However, the in-plane $\kappa_l$ from classical molecular dynamics (MD) calculations reported in the Ref. [19] seems to overcome this slightly, but the reported value still underestimates the experimental value by 15-20%. As the lattice parameters of ABaX compounds are less anisotropic, the same problem might not be true for these cases. However, the calculated $\kappa_l$ might still contain large uncertainties.

Now let's back to the electronic structure of these compounds as shown in Fig. S5. The band structure of $Bi_2Te_3$ has six extrema of both lowest conduction (LCB) and the highest valence (HVB) bands, which is consistent with the experiment [20]. Unlike the band structure of it reported by using PBE functional in Ref. [21]., the conduction band minima (CBM) and valence band maxima lie at different momentum point, i.e., indirect bandgap, which fairly agrees with the measured data from optical absorption and transmittance at room temperature [22]. Furthermore, the computed bandgap of it is in excellent agreement with the experimental gap [22]. In contrast to $Bi_2Te_3$, the LCB and HVB of NaBaBi have only two extrema and LCB of LiBaSb has three extrema.

Table II. Calculated the electronic parameters of the three compounds.

| Parameters | $Bi_2Te_3$ | | NaBaBi | LiBaSb |
| --- | --- | --- | --- | --- |
| | mBJ+SOC | Exp. | | |
| $E_g$ (eV) | 0.124 | 0.13 [22] | 0.273 | 0.8 |
| $m_h^*(m_0)$ | 1.27 | 1.26 [23] | 1.315 | 0.59 |
| $m_e^*(m_0)$ | 0.94 | 1.06 [23] | 0.83 | 0.97 |

The rotation and inversion symmetries give rise to six extrema of both LCB and HVB in $Bi_2Te_3$, while the presence of rotation symmetry only in NaBaBi causes two extrema, and rotation symmetry and only one inversion symmetry lead to three extrema of LCB in LiBaSb.

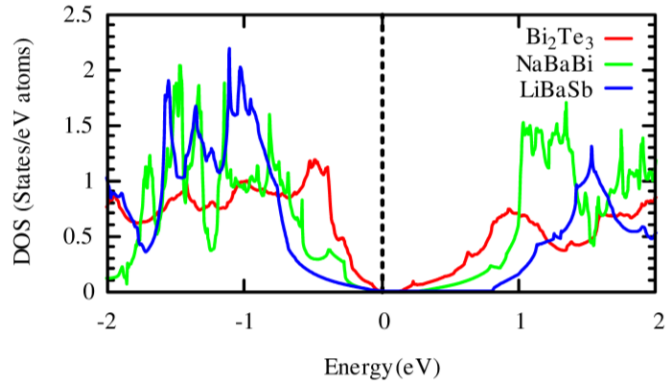

Fig. 3. The total density of states (DOS) per atoms of three compounds including spin-orbit coupling effect. The dash lines at zero energy represent the Fermi level.

From the projected density of states (as shown in Fig. S5), the HVBs arise mainly from Te-p states, while Bi-p states have a dominant contribution to the LCBs. In the case of NaBaBi and LiBaSb, the LCBs are contributed dominantly from Na-s and Li-s. Unlike in $Bi_2Te_3$, HVBs of NaBaBi and LiBaSb arise mainly from Bi-p and Sb-p states, respectively. The p-p hybridization in $Bi_2Te_3$ leads to a higher density of states around the Fermi level, while s-p hybridization in NaBaBi and LiBaSb causes a significantly lower density of states near the band edges. Moreover, the p-p hybridization in $Bi_2Te_3$ causes slightly non-parabolic CBM and VBM, while the s-p hybridization induces highly non-parabolic CBM and VBM, as shown in Fig. S6. However, the non-parabolicity of CBM and VBM in NaBaBi is more prominent than that in LiBaSb. Highly non-parabolic CBM and VBM favor high thermopower, low electronic thermal conductivity, and hence, a high figure of merit.

The calculated effective masses of these compounds from second-order polynomial fitting are listed in Table II. The present m* values of $Bi_2Te_3$ are in good agreement with the experiment [23]. The effective mass of holes of NaBaBi (LiBaSb) is slightly (significantly lower) higher than that of $Bi_2Te_3$, while the effective mass of electrons shows the opposite trend.

Fig. S7. shows the energy-dependent anisotropic lifetime of three compounds. Both LCB and HVB of ABaX have a longer lifetime, due to the relatively low density of states around Fermi level, as compared to that of $Bi_2Te_3$. The lifetime in all cases sharply falls as the carrier density rises.

The Thermopower of these compounds shows the same trend, as shown in Fig. S8. The first two compounds exhibit highly anisotropic thermopower(S), while n-type LiBaSb shows isotropic S. The increase of S of $Bi_2Te_3$ with temperature after reaching a certain carrier concentration level indicates its extrinsic behavior. NaBaBi shows this type of trend above 400K, but LiBaSb does not (within the studied temperature range). The variation of S in $Bi_2Te_3$ with temperature is more rapid than that in the other two compounds, suggesting the strong temperature dependency of effective mass. The S of ABaX is much higher, especially for n-type LiBaSb due to slightly heavier mass and wider bandgap. Although the $m_e^*$ of NaBaBi is lighter than that of $Bi_2Te_3$, almost two times wider bandgap, and highly non-parabolic bands induce higher S in NaBaBi. The non-linearity in S is due to the effect of degeneracy and mixed conduction. As the LCB and HVM of ABaX are non-degenerate bands, the S of ABaX shows linear behavior except NaBaBi at 500K, which might be due to the bipolar effect. The electrical conductivities ($\sigma$) of these compounds are shown in Fig. S9. At higher carrier density, the sig is significantly large due to mixed conduction. Overall, the $\sigma$ of $Bi_2Te_3$ and NaBaBi shows a similar trend. Although cross-plane $\sigma$ of n-type NaBaBi is comparable to that in-plane $\sigma$ of n-type $Bi_2Te_3$, the in-plane $\sigma$ above room temperature is slightly low. The $\sigma$ of $Bi_2Te_3$ exhibits slow variations with the temperature at low carrier density. Although holes effective mass of NaBaBi is heavier than that of $Bi_2Te_3$, the p-type cross-plane $\sigma$ of NaBaBi at 300K is exceptionally large due to highly non-parabolic VBM. The $\sigma$ of n-type LiBaSb is much smaller than that of the former two compounds due to its wider bandgap and heavier effective mass, but $\sigma$ of p-type LiBaSb is comparable to that of the formers. The high electrical conductivity of p-type LiBaSb is induced by the lighter effective mass of holes.

Due to the larger thermopower and comparable electrical conductivity of NaBaBi, its cross-plane power factor (PF) is almost two times higher than the in-plane PF of $Bi_2Te_3$, as shown in Fig. S10. Interestingly, the in-plane PF of NaBaBi is almost identical to the cross-plane PF of $Bi_2Te_3$. Although the PF of n-type LiBaSb much lower than that of $Bi_2Te_3$, the p-type LiBaSb has comparable PF due to its high electrical conductivity induced by light effective mass. In all cases, the room temperature PF is much higher than that at other higher temperatures, due to larger thermopower at 300K. The PF is maximum within 100-250K in all cases except p-type LiBaSb, as shown in Fig.4 (isotropic PF) and Fig.S11 (anisotropic PF).

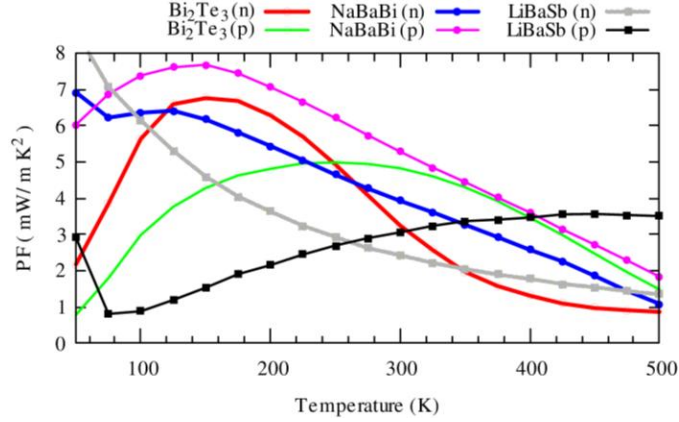

Fig. 4. The temperature-dependent isotropic power factor of three compounds under consideration. The temperature-dependent TE properties considered at the carrier concentration values are listed in Table III.

Fig. S12 shows the carrier concentration-dependent electronic part of the anisotropic thermal conductivity ($\kappa_e$) at three consecutive temperatures. Below $10^{19}$ $cm^{-3}$ carrier density, the $\kappa_e$ of ABaX is significantly lower than that of $Bi_2Te_3$, due to the relatively low density of states around the Fermi level of ABaX compounds. But above this limit, $\kappa_e$ of ABaX (except n-type LiBaSb) is higher than that of $Bi_2Te_3$. This may be due to the effect of mixed conduction. This feature of $\kappa_e$ in ABaX will significantly enhance its TE performance.

Table III. Calculated room temperature isotropic TE parameters of $Bi_2Te_3$ and ABaX. The experimental data for n- and p-type $Bi_2Te_3$ are taken from Ref. [24] and Ref. [25], respectively.

| Compound | | n ($10^{18}$ $cm^{-3}$) | \|S\| (μV K$^{-1}$) | σ ($10^5$ $S\,m^{-1}$) | PF ($mW\,m^{-1}K^{-2}$) | κtot ($W\,m^{-1}K^{-1}$) | ZT |
|---|---|---|---|---|---|---|---|
| $Bi_2Te_3$ | n | 1.0 | 179 | 0.958 | 3.07 | 2.13 | 0.432 |
| | Exp. | 16.6 | 190 | 0.696 | 2.51 | 1.75 | 0.43 |
| | p | 10.2 | 130 | 2.78 | 4.7 | 2.18 | 0.65 |
| | Exp. | 11.0 | 219 | 0.5 | 2.4 | 2.02 | 0.36 |
| NaBaBi | n | 1.0 | 243 | 0.616 | 3.64 | 1.01 | 1.08 |
| | p | 1.1 | 218 | 1.06 | 5.04 | 1.13 | 1.34 |
| LiBaSb | n | 4.3 | 322 | 0.233 | 2.42 | 1.86 | 0.39 |
| | p | 11.0 | 94 | 3.79 | 3.35 | 3.49 | 0.29 |

The larger PF, lower ke, and almost similar lattice thermal conductivity of NaBaBi lead to a high figure of merit (ZT) as shown in Fig. S13. The cross-plane ZT of NaBaBi is almost 2.5 and 3.1 times larger for n-type and p-type respectively than the corresponding in-plane ZT of $Bi_2Te_3$. Besides, the in-plane ZT of NaBaBi is almost two times larger than the cross-plane ZT of $Bi_2Te_3$. Interestingly, the room temperature ZT of LiBaSb is almost the same (except cross-plane ZT of p-type LiBaSb, in that case, ZT is exceptionally high) to that of $Bi_2Te_3$ and the ZT of LiBaSb reaches its maximum value at 500 K( within the temperature limit under consideration). This is because of its relatively wider bandgap.

Fig. S14 shows the temperature effect on the thermoelectric performance of three compounds. The ZT of n-type $Bi_2Te_3$ is maximum below room temperature, which is consistent with the experiment. In the case of p-type, the ZT becomes a maximum above 300K. In contrast to $Bi_2Te_3$, the ZT of ABaX becomes maximum above room temperature for both types of carriers, which might be due to a relatively wider bandgap. The room temperature TE parameters of the three compounds are listed in Table III.

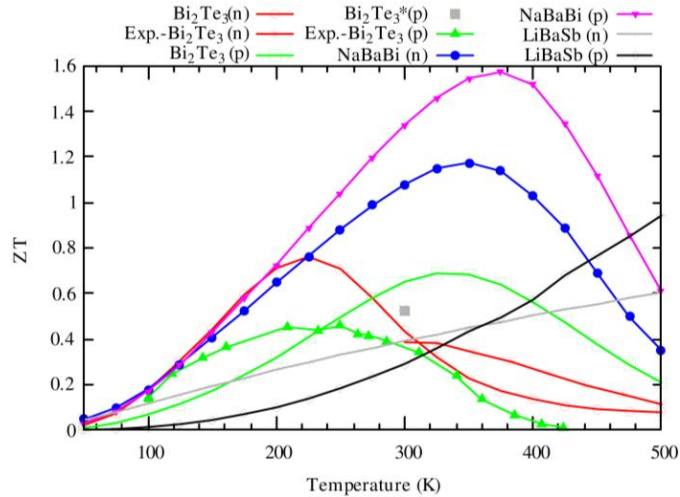

Fig. 5. Predicted isotropic thermoelectric figure of merit (ZT) at different temperatures of (a) $Bi_2Te_3$, (b) NaBaBi, (c) LiBaSb. Here I used the calculated values of lattice thermal conductivity, although the calculated in-plane lattice thermal conductivity largely underestimates the experimental value (extracted from total thermal conductivity). The square symbol presents the calculated ZT of p-type $Bi_2Te_3$.

The computed isotropic ZT of the three compounds including experimental data of $Bi_2Te_3$ is shown in Fig. 5. Although the calculated ZT of $Bi_2Te_3$ around room temperature fairly agrees with the experiment (especially for n-type carriers), it rapidly diverges both at low and high temperatures. A similar overestimation of ZT of p-type $Bi_2Te_3$ from first-principles calculations was reported in Refs. [21,26]. At these temperatures, there are might be some complex scattering mechanism involved which semi-classical Boltzmann equation. The overestimation of ZT (about 50% at 300K in the case of p-type $Bi_2Te_3$, even if the experimental $\kappa_l$ is used (about 40%))is mainly due to the severe underestimation of the in-plane lattice thermal conductivity and overestimation of the experimental electrical conductivity. If the same uncertainty is considered in the case of NaBaBi, the ZT will still much larger, about ~0.6 and ~0.8 for n- and p-type carriers at 350K. Thus, within the computational accuracy, NaBaBi is an extraordinary thermoelectric material having much better TE performance than that of $Bi_2Te_3$. On the other side, the room temperature isotropic ZT of both n- and p-type LiBaSb is much smaller than that $Bi_2Te_3$, but is much larger at 500 K, suggesting its potential TE applications above 400 K.

In summary, from first-principles calculations, I report the details of electronic and thermoelectric properties of $Bi_2Te_3$, NaBaBi, and LiBaSb. The present calculations suggest that highly non-parabolic bands and two times wider bandgap of NaBaBi compared to $Bi_2Te_3$ lead to a much higher thermopower, while almost similar effective mass and low density of states around Fermi level help to remain the electrical conductivity almost the same. Interestingly, the low DOS of NaBaBi causes much lower electronic thermal conductivity at optimum carrier concentrations compared to $Bi_2Te_3$. Furthermore, NaBaBi has similar lattice thermal conductivity due to strong anharmonicity and exhibits weak anisotropic behavior. On the other hand, the electrical conductivity LiBaSb is much lower for n-type carriers due to its wide bandgap, but comparable for p-type carrier due to its low effective mass, to that of $Bi_2Te_3$. Moreover, its higher lattice thermal conductivity and wider bandgap make it less suitable for TE application at low temperatures, but it has a strong potential for medium-range temperature applications. The extraordinary thermoelectric figure of merit of NaBaBi is expected to replace the use of $Bi_2Te_3$ and its alloys in commercial TE device applications soon.

**Supporting Information**

Extraordinary thermoelectric performance of ABaX compared to $Bi_2Te_3$

Enamul Haque

EH Solid State Physics Laboratory, Longaer, Gaffargaon-2233, Mymensingh, Bangladesh.
Email: enamul.phy15@yahoo.com, enamul@mailaps.org

**Computational Details**

In this study, I used a set of first-principles codes to calculate the different types of properties. First I performed structural relaxation by using the plane wave method as implemented in Quantum

Espresso. In the calculation, I set a very strict convergence criterion (energy convergence $10^{14}$ Ry, force 0.1 mRy/au and Pulley stress 0.1 kbar) to obtain the ground state structure. The exchange-correlation part was treated through generalized Gradient approximation (GGA) with PBEsol setting by using PAW for $Bi_2Te_3$ and ultrasoft Vanderbilt pseudopotential for ABaX. I selected a 41.52, 48.5, and 55.5 Ry cutoff energy for wavefunction, 166, 194, and 222 Ry for charge density and, $6 \times 6 \times 6$, $6 \times 6 \times 10$, $8 \times 8 \times 4$ Γ-centered k-point with Marzari-Vanderbilt smearing of width 0.03 Ry after extensive trials. Since the electron-phonon matrix calculation is very expensive, I used a 444, 444, and 442 uniform q-point grid (and $8 \times 8 \times 8, 8 \times 8 \times 12, 12 \times 12 \times 6$ k-point mesh) to reduce the computational burden. The average electron-phonon dynamical matrix was calculated by using EPA code. The numbers of energy bins used in these calculations are 10, 6, and 8 after extensive trials.

The use of relatively loose q-point might have a slight negative impact on the accuracy of the electron-phonon scattering matrix. To check this impact, the calculations of the average e-ph matrix were repeated by using a moving least square method (MLS) electron-phonon averaged approximation with 30 energy bins for each compound, which is less sensitive (even it allows to use 222 q-point grid without the loss of significant accuracy) to the q-point grid and found negligible impact. The matrix was then fed into slightly modified BoltzTraP code to calculate transport coefficients. This code uses the semiclassical Boltzmann transport theory and thus, requires accurate electronic structure calculations. To calculate accurate electronic structure, I used Tran-Blah modified Becke-Johnson potential, as implemented in wien2k, a full-potential linearized augmented plane wave method based code. To proceed this calculation, I first minimized the atomic forces again in Wien2k by using PBEsol, with the same k-point, plane-wave cutoff $RK_{max}=7$, valence and core states separation energy -10.0 and -6.0 Ry for $Bi_2Te_3$ and ABaX, muffin tin sphere radii 1.96 Bi and Te, 2.1 and 2.3 for Na and Ba/Bi, 2.19 and 2.5 Bohr for Li and Ba/Sb, respectively. I then performed the electronic structure calculation by using a denser $32 \times 32 \times 32$, $30 \times 30 \times 43$, and $44 \times 44 \times 21$ non-shifted k-point mesh to obtain energy eigenvalues. In the electronic structure and transport calculations, I included the spin-orbit coupling (SOC) effect explicitly by performing fully-relativistic calculations.

I calculated lattice thermal conductivity ($\kappa_l$) by using 221, 112, and 221 supercells for $Bi_2Te_3$, NaBaBi, and LiBaSb, respectively, as implemented in phono3py. To calculate, third-order IFCs and second-order force constants, the force calculations were performed in QE (with the same setting as before) for each displacement. Note this type of calculation is very expensive and spin-orbit coupling has little bit effect on the lattice thermal conductivity, so this effect was not included in the force calculations. After force calculations, $\kappa_l$ was obtained by solving the linear Boltzmann phonon equation with $16 \times 16 \times 16$ q-point. Note that q-point convergence was also checked by using a set of different q-points.

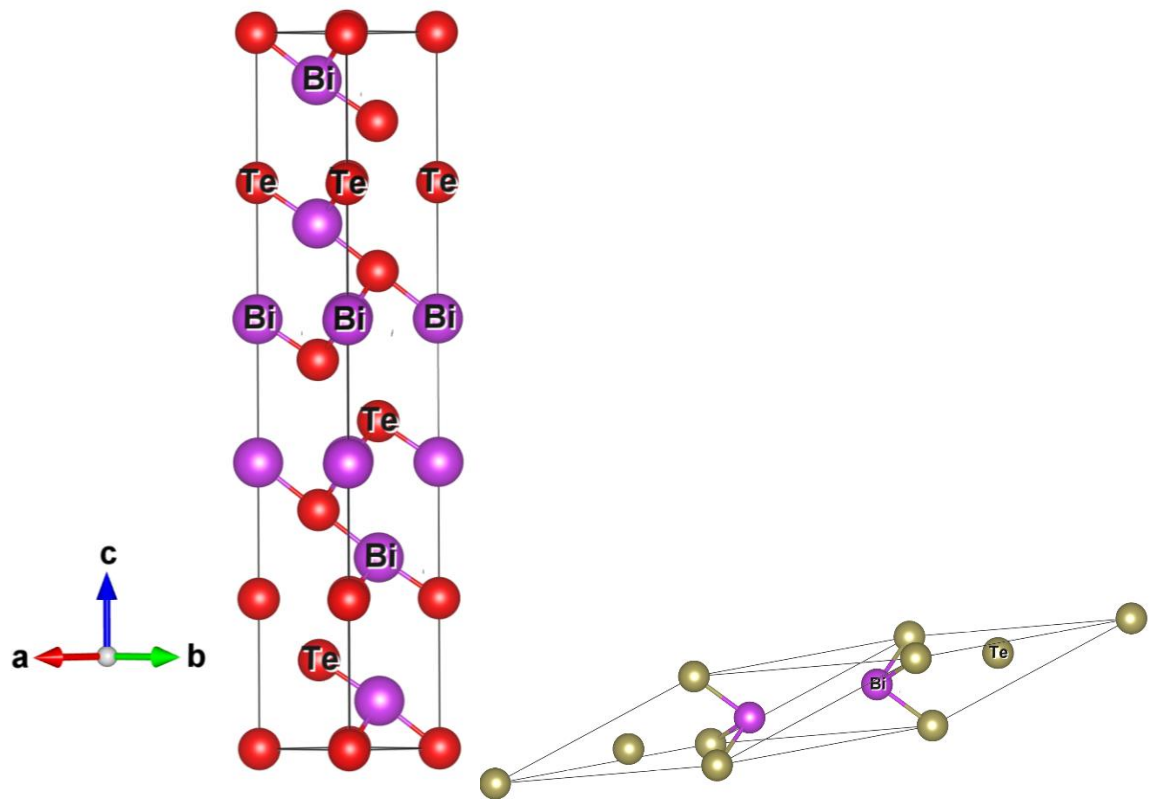
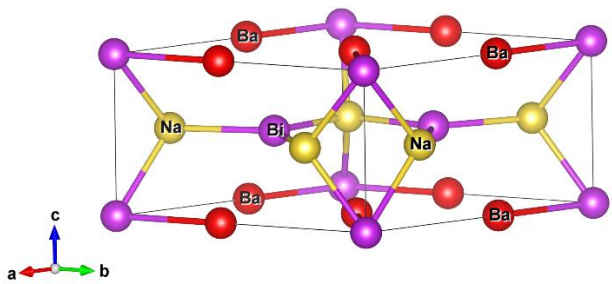
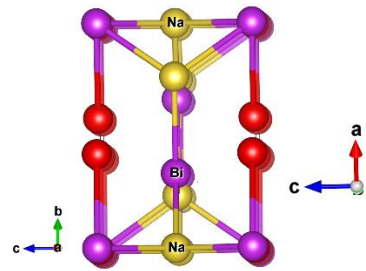
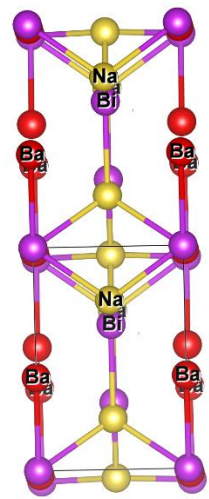

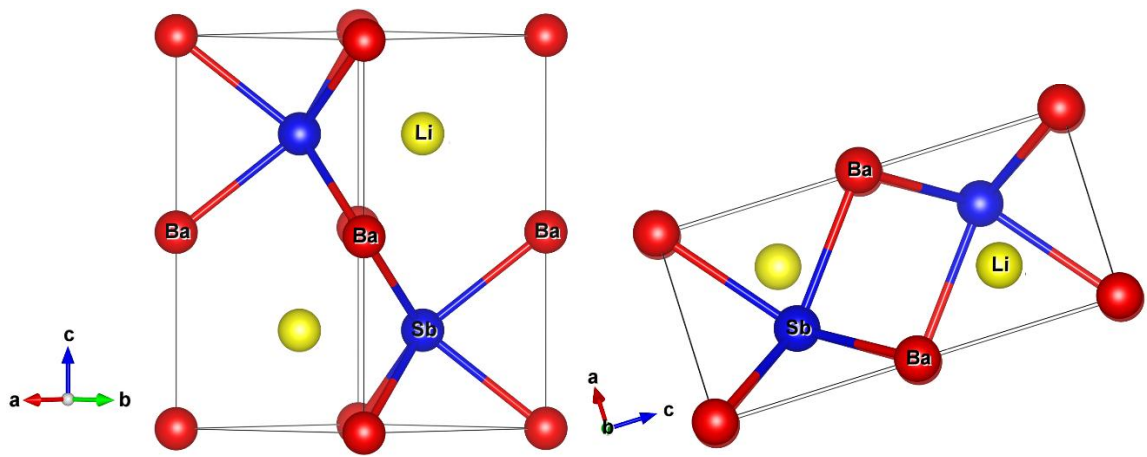

Fig. S1. Ground state crystal structures. The top panel shows the conventional and primitive unit cells of Bi$_2$Te$_3$; the middle panel shows the unit cell of NaBaBi with two side views, and the bottom panel shows the unit cell of LiBaSb with one side view.

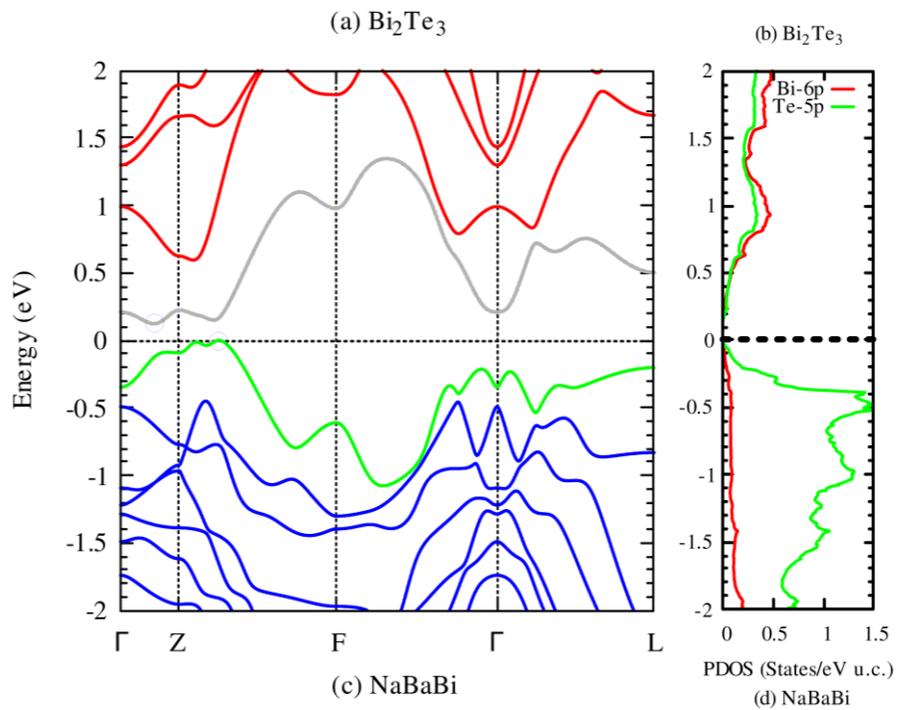

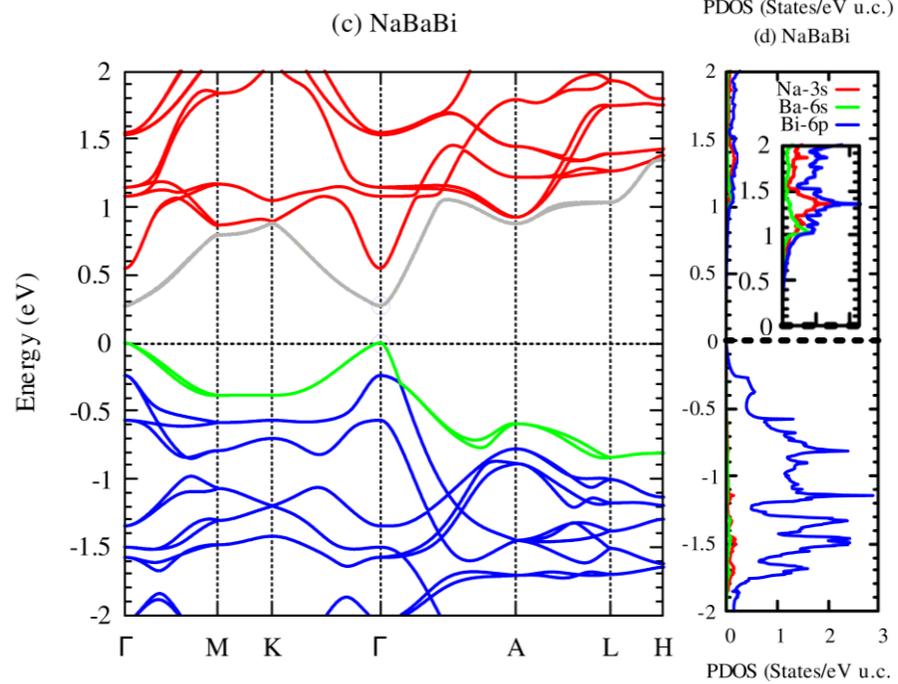

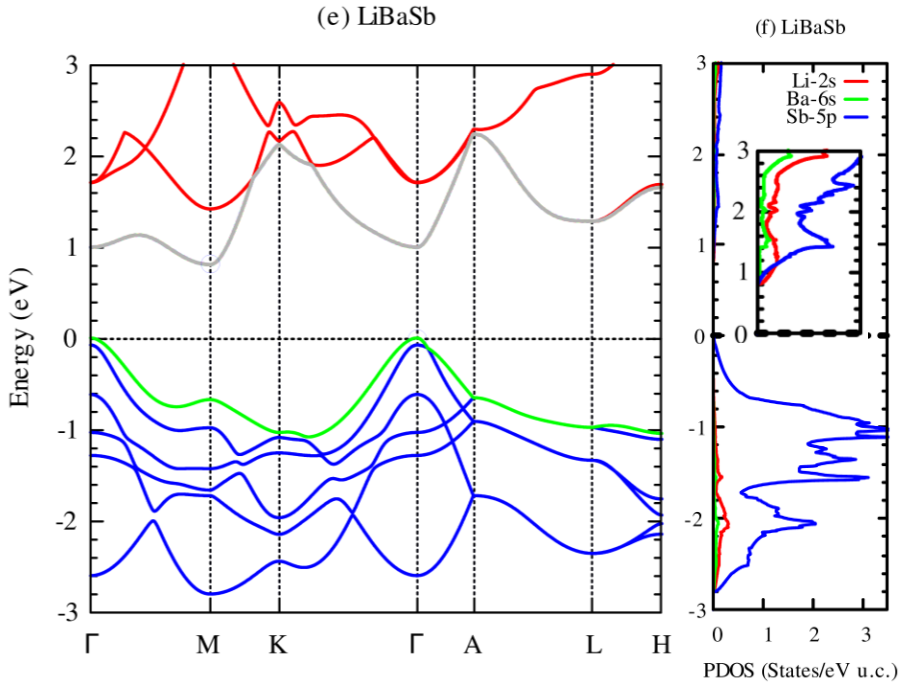

Fig. S2. Electronic dispersion relations and projected density of states of $Bi_2Te_3$ (a-b), NaBaBi (c-d) and LiBaSb (e-f). The open circle indicates the conduction band minima and valence band maxima. The dashed lines at zero energy represent the Fermi level.

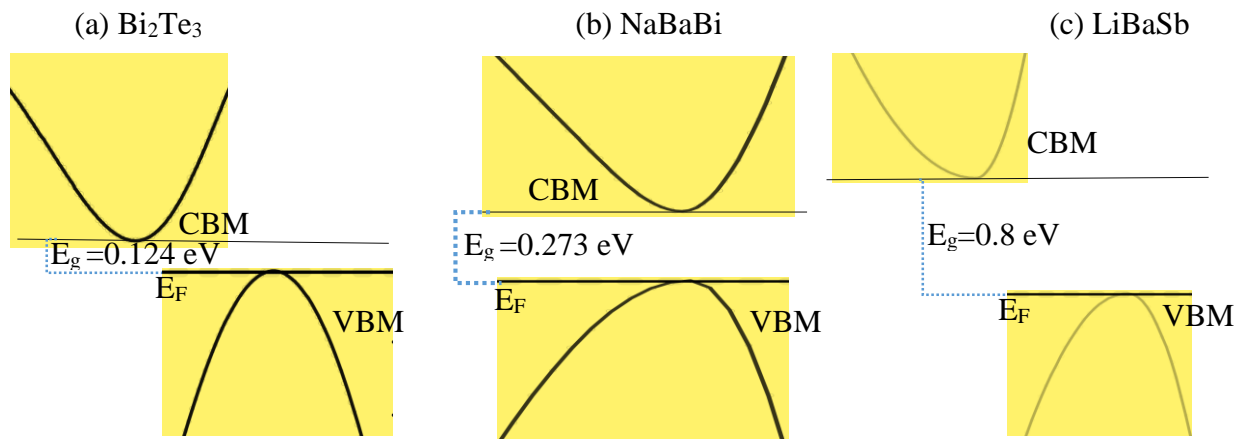

Fig. S3. Non parabolic nature of CBM and VBM of $Bi_2Te_3$ (a), NaBaBi (b), and LiBaSb (c).

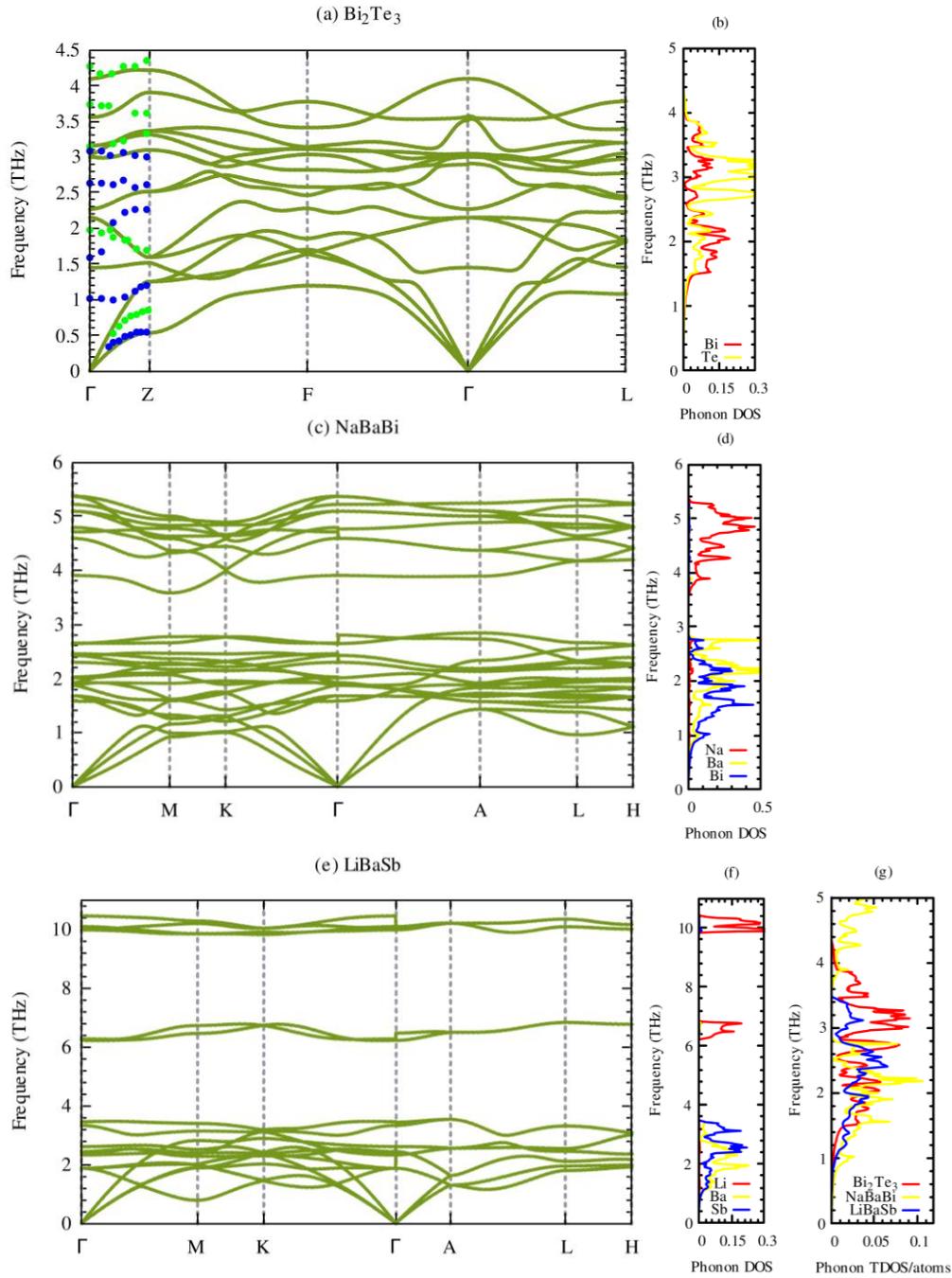

Fig. S4. Phonon dispersion relations over the high symmetry points of Brillouin zone (BZ) and partial phonon density of states of $Bi_2Te_3$ (a-b), NaBaBi (c-d), and LiBaSb (e-f). Total phonon density of states per atom of these three compounds (g). The colored circles in (a) represent the experimental phonon dispersion of $Bi_2Te_3$ taken from Ref. [27].

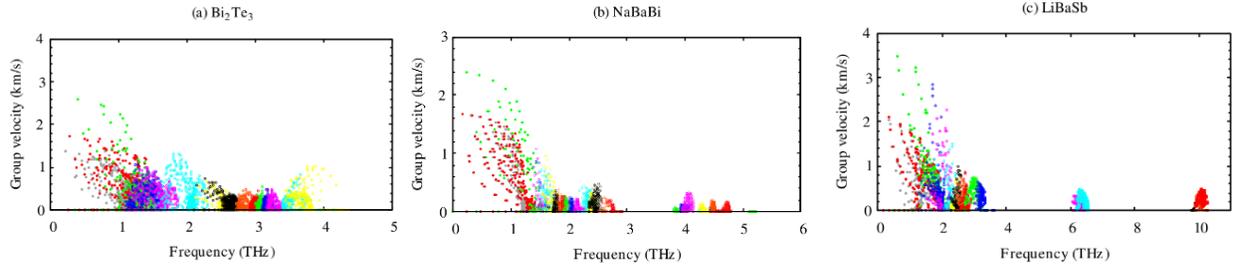

Fig. S5. Computed phonon group velocity of (a) $Bi_2Te_3$, (b) NaBaBi, and (c) LiBaSb from second-order harmonic IFCs.

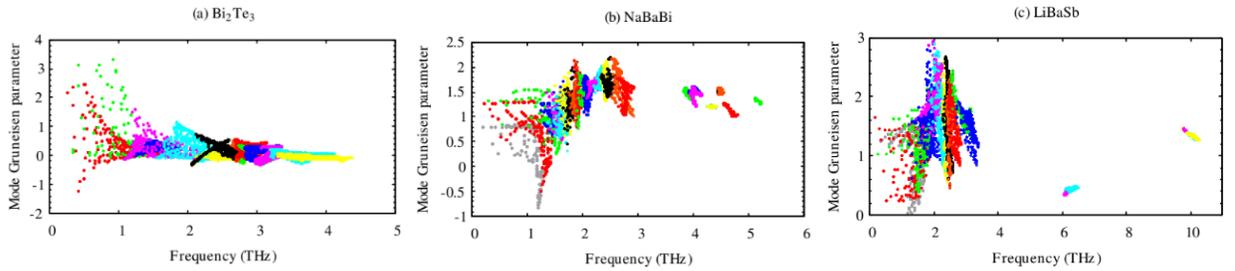

Fig. S6. Mode Gruneisen parameter of (a) $Bi_2Te_3$, (b) NaBaBi, and (c) LiBaSb calculated from anharmonic IFCs.

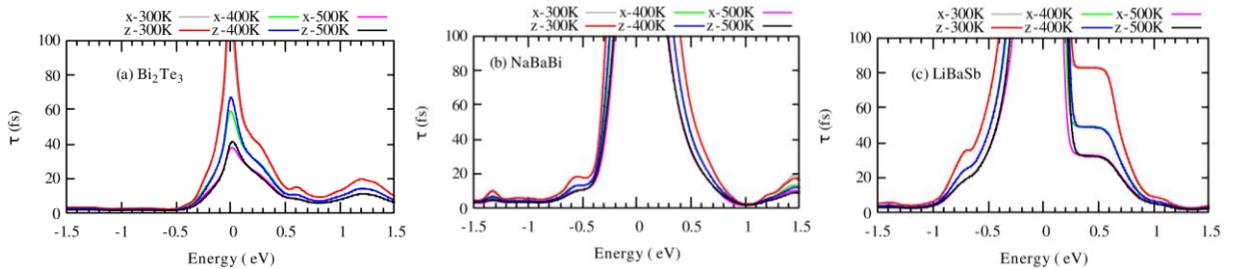

Fig. S7. Energy-dependent anisotropic carrier lifetime ($\tau$) at three consecutive temperatures of (a) $Bi_2Te_3$, (b) NaBaBi, and (c) LiBaSb. The Fermi level was set to zero.

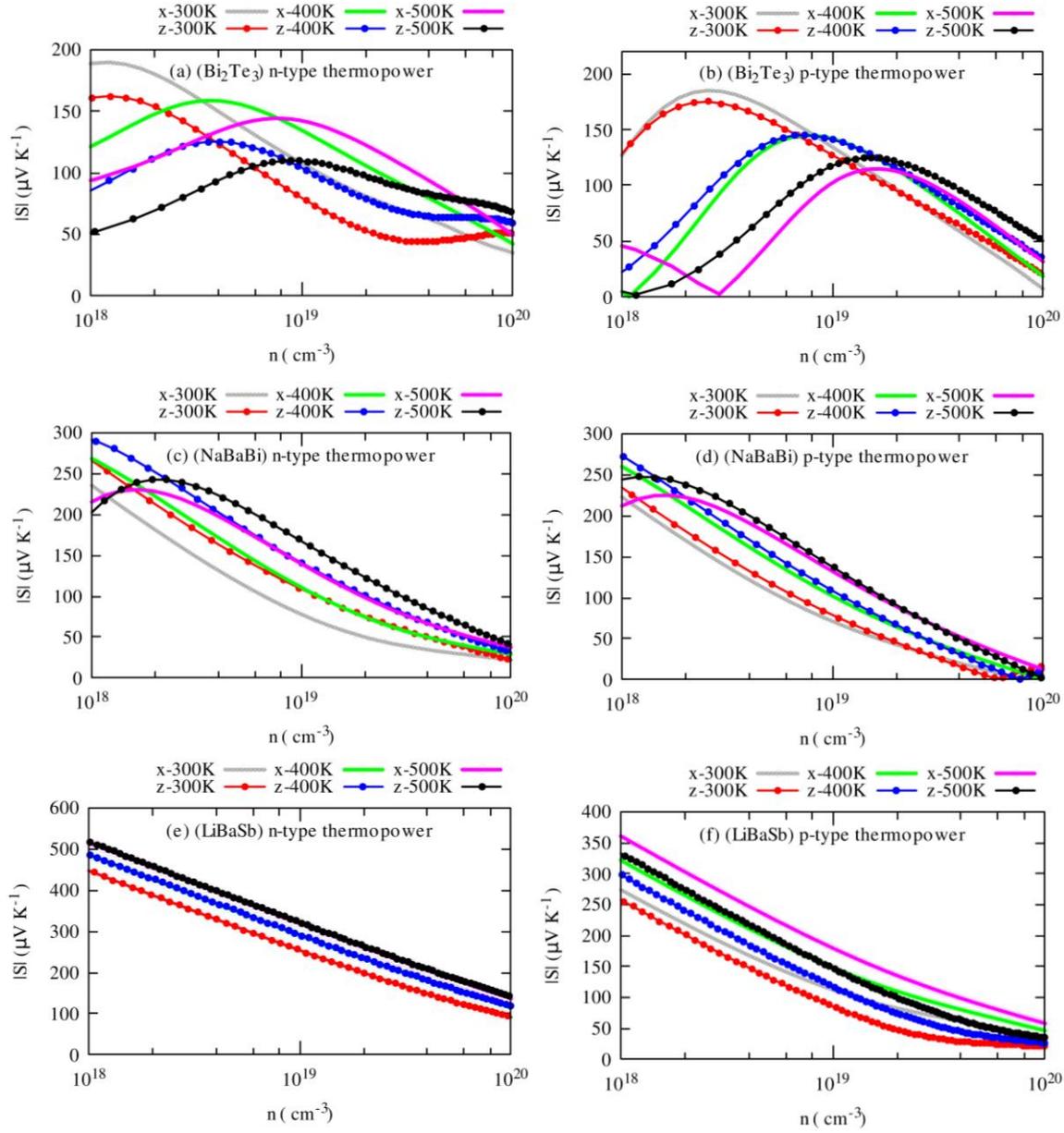

Fig. S8. Carrier concentration dependent absolute values of in-plane (x) and cross-plane (z) thermopower of (a-b) $Bi_2Te_3$, (c-d) NaBaBi, and (e-f) LiBaSb for n- and p-type carriers at three consecutive temperatures.

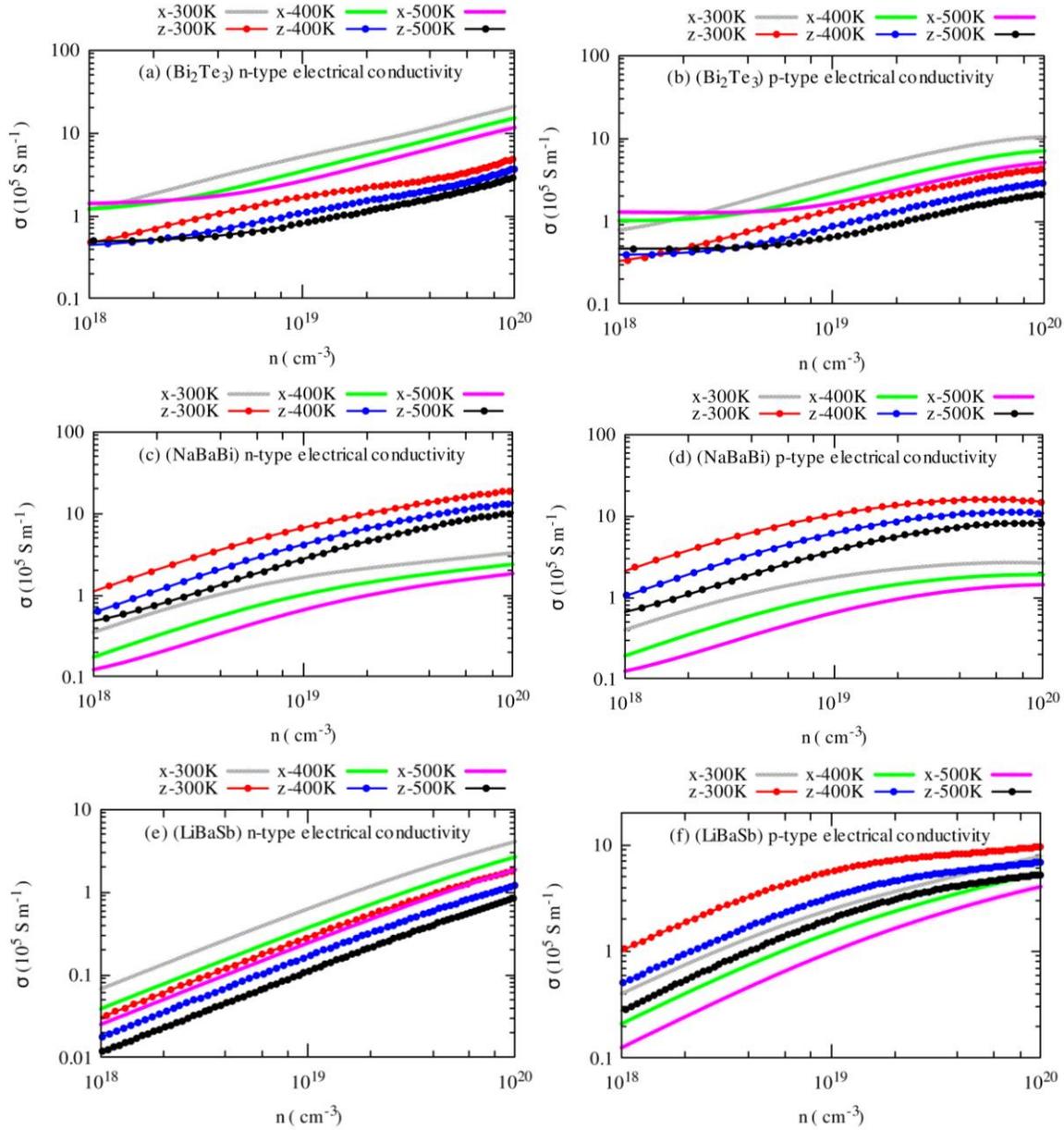

Fig. S9. Carrier concentration dependent in-plane (x) and cross-plane (z) electrical conductivity of (a-b) $Bi_2Te_3$, (c-d) NaBaBi, and (e-f) LiBaSb for n- and p-type carriers at three consecutive temperatures.

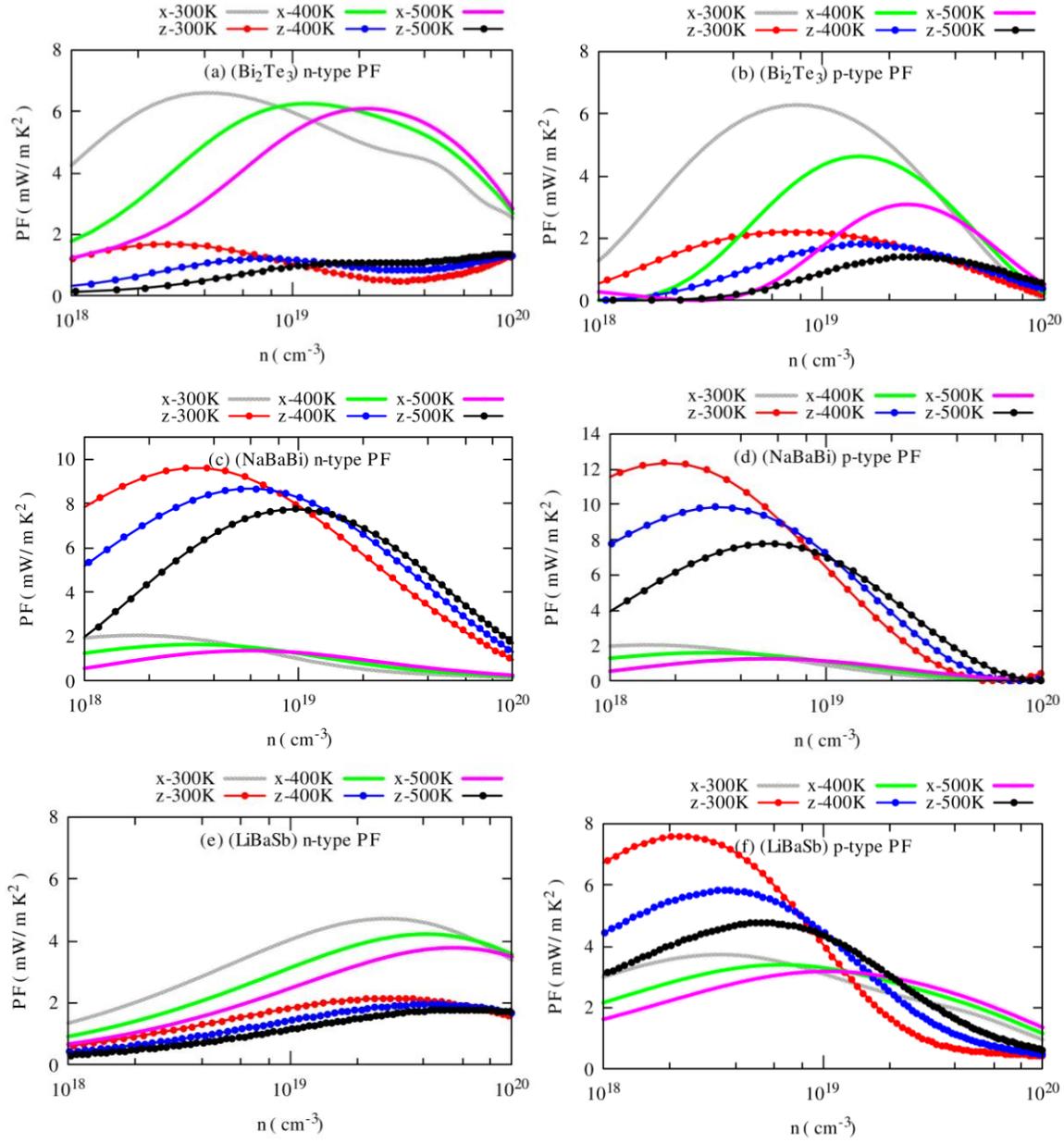

Fig. S10. Calculated in-plane (x) and cross-plane (z) power factor as a function of carrier concentration of (a-b) $Bi_2Te_3$, (c-d) NaBaBi, and (e-f) LiBaSb for n- and p-type carriers at three consecutive temperatures.

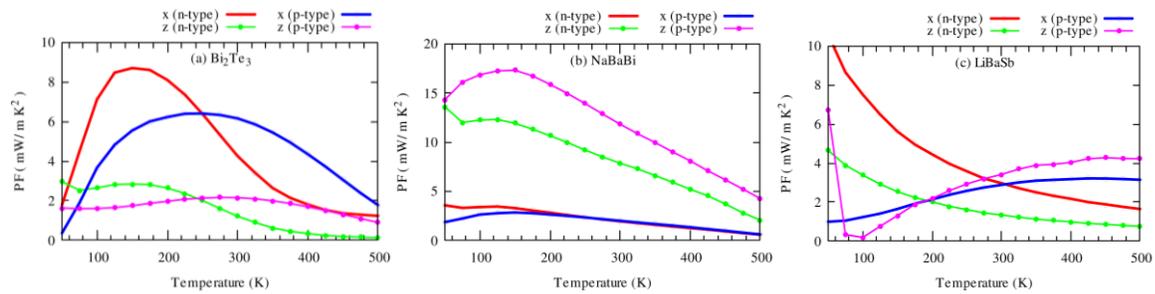

Fig. S11. Temperature dependent anisotropic power factor of (a) $Bi_2Te_3$, (b) NaBaBi, and (c) LiBaSb.

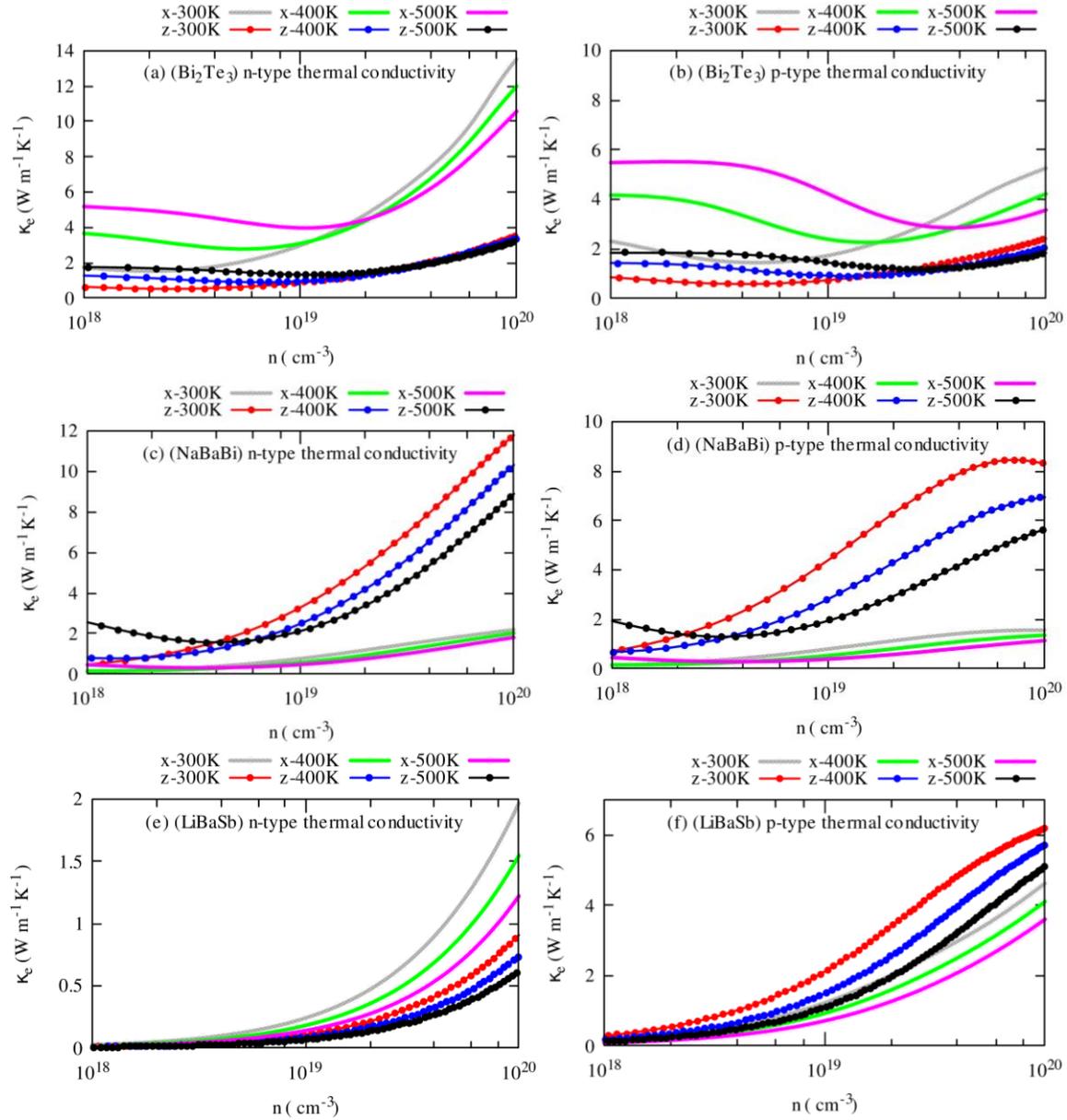

Fig. S12. Carrier concentration dependency of in-plane (x) and cross-plane (z) electronic part of the thermal conductivity of (a-b) $Bi_2Te_3$, (c-d) NaBaBi, and (e-f) LiBaSb for n- and p-type carriers at three consecutive temperatures.

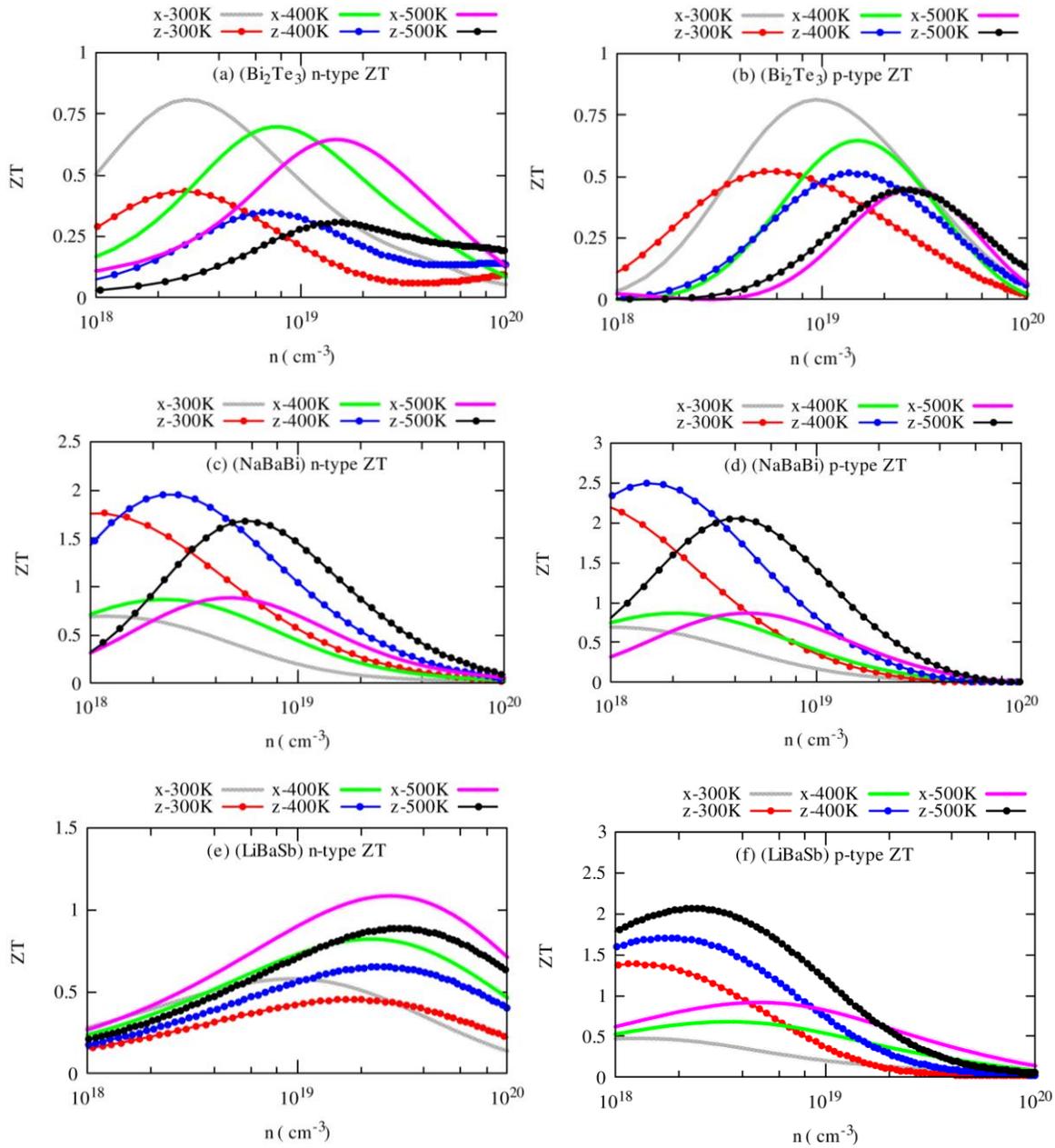

Fig. S13. Predicted anisotropic thermoelectric figure of merit as a function of a carrier concentration of (a-b) $Bi_2Te_3$, (c-d) NaBaBi, and (e-f) LiBaSb for n- and p-type carriers at three consecutive temperatures.

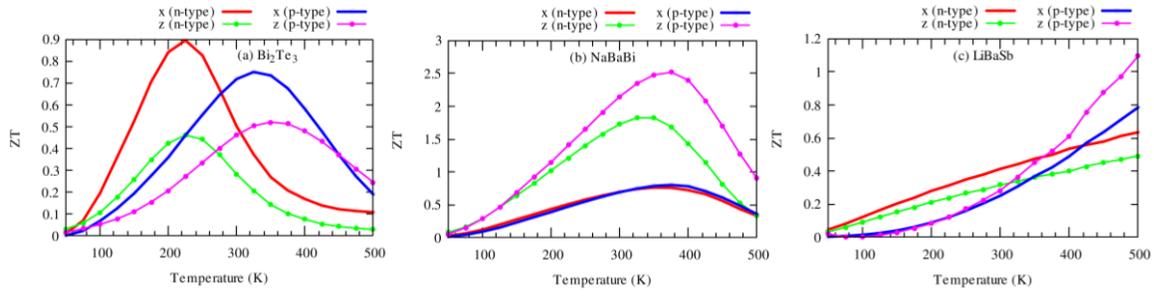

Fig. S14. Computed temperature-dependent anisotropic figure of merit (ZT) of (a) $Bi_2Te_3$, (b) NaBaBi, and (c) LiBaSb.